\documentclass[%
 aip, 
 jmp,%
 amsmath,amssymb,
 fleqn,
 preprint,%
]{revtex4-1}

\usepackage{textcomp}
\usepackage{graphicx}
\usepackage{dcolumn}
\usepackage{bm}

\begin{document}

\title {3-D particle-in-cell simulations for quasi-phase matched direct laser electron acceleration in density-modulated plasma waveguides}
\author{M.-W. Lin}
\affiliation{Department of Mechanical and Nuclear Engineering,\\
The Pennsylvania State University, University Park, PA 16802, USA}

\author{Y.-L. Liu}
\affiliation{Department of Physics, National Central University, Jhongli 32001, Taiwan}

\author{S.-H. Chen}
\email{chensh@ncu.edu.tw}
\affiliation{Department of Physics, National Central University, Jhongli 32001, Taiwan}

\author{I. Jovanovic}
\affiliation{Department of Mechanical and Nuclear Engineering,\\
The Pennsylvania State University, University Park, PA 16802, USA}


\begin{abstract}

Quasi-phase matched direct laser acceleration (DLA) of electrons can be realized with guided, radially polarized laser pulses in density-modulated plasma waveguides. A 3-D particle-in-cell model has been developed to describe the interactions among the laser field, injected electrons, and the background plasma in the DLA process. Simulations have been conducted to study the scheme in which seed electron bunches with moderate energies are injected into a plasma waveguide and the DLA is performed by use of relatively low-power (0.5-2 TW) laser pulses. Selected bunch injection delays with respect to the laser pulse, bunch lengths, and bunch transverse sizes have been studied in a series of simulations of DLA in a plasma waveguide. The results show that the injection delay is important for controlling the final transverse properties of short electron bunches, but it also affects the final energy gain. With a long injected bunch length, the enhanced ion-focusing force helps to collimate the electrons and a relatively small final emittance can be obtained. DLA efficiency is reduced when a bunch with a greater transverse size is injected; in addition, micro-bunching is clearly observed due to the focusing and defocusing of electrons by the radially directed Lorentz force. DLA should be performed with a moderate laser power to maintain favorable bunch transverse properties, while the waveguide length can be extended to obtain a higher maximum energy gain, with the commensurate increase of laser pulse duration and energy.    
           
\end{abstract}

\maketitle

\section{\label{sec_Intro}Introduction}
The limitations on the accelerating field amplitude in radio-frequency accelerators motivate the development of alternative accelerator technologies with much greater acceleration gradients. Laser wakefield acceleration (LWFA)~\cite{Tajima1979PRL,Esarey2009RMP} is one such alternative method, which relies on intense laser pulses to excite plasma waves and utilizes the high electric fields in a plasma for electron acceleration. LWFA has been demonstrated in numerous experiments and is a very active research area at present time. Electron bunches with energies up to GeV level have been produced by using laser pulses with peak powers ranging from tens of TW to PW.~\cite{Leemans2006NPhys,Wang2013NCom} Alternatively, it has been proposed to use the powerful electromagnetic field of a laser pulse to directly accelerate electrons without an intervening medium~\cite{Malka1997PRL}. Such schemes belong to the class of direct laser acceleration (DLA) methods. However, if the phase velocity of the optical field $v_{p}$ is greater than the electron velocity $v_{e}$, the Woodward-Lawson theorem states that zero net energy gain is produced over an infinite acceleration distance.~\cite{LWT,Esarey1995PRE} For DLA methods operating in photonic bandgap (PBG) materials,~\cite{Lin2001PRSTAB,Mizrahi2004PRE} the working principle is to induce a resonant wave in the center channel (with a width on the scale of optical wavelength) which propagates with a phase velocity $v_p \leq c$ to accelerate co-propagating electrons. A grating-based DLA structure results in electrons experiencing a greater field amplitude in the acceleration phase than in the deceleration phase, such that a net energy gain is accumulated.~\cite{Peralta2013Nature} In a similar manner, guiding a radially polarized laser pulse~\cite{Salamin2006NJF} in a plasma waveguide has been proposed for realizing DLA~\cite{Serafim2000IEETPS} in which the co-propagating electrons are accelerated by the axial electric field of the laser pulse. Because a plasma waveguide extends the acceleration distance, the required laser power can be considerably reduced from that in unguided DLA,~\cite{Unguid_DLA} where hundreds of TW of laser power would be needed to obtain electron energies in the range of hundreds of MeV.

  A significant challenge for realizing DLA in a plasma waveguide is identifying a phase matching mechanism between the accelerated electrons and the laser field, which propagates at a superluminal phase velocity $v_{p}>c$ due to the presence of the plasma.~\cite{Serafim2000IEETPS} Analogous to the ``slow-wave'' structures extensively used for RF waves, a periodic density structure in a plasma waveguide expands the laser axial field into several harmonics, for which quasi-phase-matching (QPM) of DLA has been proposed and well studied.~\cite{DLA_OSWS,Yoon2012PRSTAB} The results of prior simulations show that the harmonic axial field component, having a subluminal phase velocity $v_{p}<c$, can effectively accelerate electrons along the waveguide when a proper density modulation period for the QPM condition, including any necessary density ramping,~\cite{Yoon2012PRSTAB} has been prepared. The QPM process of DLA can also be understood in an alternative picture, by breaking the energy gain symmetry between the acceleration and deceleration phases, similar to the processes in quasi-phase-matched relativistic harmonic generation.~\cite{Kuo2007PRL,Liu2008POP} Since the electrons fall out of a phase by $\pi$ with respect to the co-propagating laser field over a dephasing length $L_{d}$, QPM of DLA can be achieved by preparing an axial density modulation in a plasma waveguide with alternating low- and high-density regions~\cite{Lin2012POP}, taking advantage of the plasma density-dependent dephasing length $L_d$. Electrons injected at proper phases for QPM of DLA gain more energy in the longer, low-density regions than they lose in the shorter, high-density regions; thus a net energy gain is accumulated. A density-modulated plasma waveguide that can support this mechanism of net acceleration can be fabricated via the optical laser machining technique,~\cite{Lin2006POP,Hung2012POP} which represents a modification of the igniter-heater scheme.~\cite{Volfbeyn1999POP} 
 
  For DLA realized in a plasma waveguide, the electron bunch interacts with the electromagnetic field of a co-propagating laser pulse, but also with the electric field originating from the plasma density perturbation driven by the bunch charge and the laser ponderomotive force. The contribution to the total electric field experienced by electrons, originating from the plasma response in DLA, has not been considered in the previous test particle model~\cite{Lin2012POP} due to the complexity of simulating nonlinear plasma interactions in that simulation approach. To improve the fidelity of the DLA simulation, particle-in-cell (PIC) simulation can be used for more detailed studies of DLA with closed-loop solutions for the variation of the electromagnetic field and the dynamics of charged particles for the injected bunch and the background plasma. PIC simulations offer a significantly improved understanding of the electron bunch properties during the DLA process. For example, the electron bunch collimation can be improved by using a higher density for the injected electron bunch, as has been shown in a recent 2-D PIC study.~\cite{Yoon2012PRSTAB} 
 
 In this work, a 3-D PIC model has been developed and used to simulate QPM of DLA in density-modulated plasma waveguides. Sharp axial periodic structures with alternating waveguide and neutral gas regions are defined in the simulation to reproduce the properties of a laser-machined plasma waveguide.~\cite{Lin2006POP} The waveguide regions contain a radially increasing plasma density to provide the guiding force to the laser pulse and serve as the low-density regions needed to support the QPM mechanism. A higher atom density with a uniform density profile is assigned to the neutral gas regions. The model takes into account the optical-field-ionization (OFI) of the gas atoms by the laser pulse in the neutral gas regions. After the OFI of neutral gas atoms by the laser front foot, most of the drive laser field and the injected bunch electrons experience alternating low and high plasma density regions along the propagation distance. Thus, the axial QPM condition for DLA is determined by the lengths of the waveguide and neutral gas regions. This model also helps to identify if the laser pulse energy depletion in plasmas and/or pulse defocusing in the neutral gas regions can possibly render DLA ineffective, as discussed previously.~\cite{Lin2012POP} The presented DLA test particle analysis has shown that larger axial and radial acceleration regions exist for electrons injected with a higher initial energy. With a low initial energy of a few MeV, electrons injected in a suboptimal phase or at a greater radial position cannot be accelerated to the required energies for consistently meeting the QPM condition along the propagation distance. Therefore, QPM of DLA is predicted to have higher acceleration efficiency if the electron bunches can be pre-accelerated to tens of MeV before injection. Several LWFA experiments have demonstrated the production of electron bunches in the energy range of 20-50 MeV by using laser pulses with peak power of a few TW.~\cite{LWFA_Eng} Therefore, we use the PIC simulations to study the scheme in which an electron bunch from a LWFA is injected into the plasma waveguide for the second-stage of QPM of DLA to higher energies.  

The injected bunch electrons will simultaneously experience multiple forces throughout the DLA process. In addition to being driven directly by the laser field, the injected electrons for DLA also experience the nonlinear laser ponderomotive force and the electrostatic force from the resulting density variation of the background plasma electrons. The hollow intensity distribution of the laser radial field is a particular case where the ponderomotive force pushes the background plasma electrons to concentrate in the center, which in turn produces a radial electrostatic force that can defocus the injected electron bunch.~\cite{Yoon2012PRSTAB} The electron bunch also expands because of its finite emittance. Analogous to the Rayleigh length of a laser beam, the minimum $\beta$-function $\beta^{*}=\gamma \sigma_{y}^{2}/\epsilon_{N}$ characterizes the beam size variation along the propagation, where $\gamma$ is the Lorentz factor of the bunch, $\sigma$ is the root-mean-square (RMS) bunch size and $\epsilon_{N}$ is the normalized emittance.~\cite{Rosenzweig2003Book} In contrast, the ponderomotive force of the laser radial field also provides a confinement force for the electron bunch, similar to the effect of pushing the plasma electrons to concentrate in the center. The electron bunch will experience the maximum focusing force when it becomes synchronized with the peak of the laser pulse envelope. Moreover, the injected electron bunch interacts with the plasma electrons and drives a plasma wave.~\cite{Esarey2009RMP} In an overdense plasma ($n_{b} \ll n_{p0}$), the electron bunch is focused by the induced wake-field of a linear plasma wave.~\cite{Chen1987IEEEPS,Rosenzweig1989PRD,Rosenzweig1990PFB} When $n_{b}>n_{p0}$ of a underdense plasma condition, a large portion of background electrons is ejected by the bunch head, leaving an ion channel in a nonlinear plasma wave that can exerts a focusing force to the bunch correspondingly.~\cite{Su1990PRA,Rosenzweig1991PRA,Barov1994PRE} 
    
The goal of this work is to understand how the initial parameters of the injected bunch can be chosen to optimize the DLA. Selected time delays (with respect to the laser pulse), bunch lengths and bunch sizes are assigned to the injected electrons in a series of simulations. We analyse the energy spectrum, trace space, emittance and density (or electron particle) distribution of the injected bunch to understand the variation of bunch properties throughout the DLA process and how they relate to the initial conditions. In Sec.~\ref{sec_model}, a detailed description for the 3-D PIC model is introduced, along with verification for the simulated dephasing length $L_{d}$ and the review of the initial bunch properties. Simulation results and discussions of DLA performance are presented in Sec.~\ref{sec_Results}. DLA of short electron bunches (on the order of few fs long) injected at selected time delays with respect to the laser pulse are studied first. The results show that the final bunch emittance is highly correlated with the bunch injection delay, which is explained by the density perturbation induced by the laser pulse and the radial Lorentz force that drives the radial dynamics of off-axis electrons. Next, selected bunch lengths and bunch sizes are assigned to the injected electron bunches in a series of simulations. Mechanisms for the bunch density modulation throughout the DLA process are discussed. Simulations are then used to study the properties of accelerated electron bunches as the laser power and the plasma waveguide length are increased. Finally, a discussion and the summary of this work are provided in Sec.~\ref{sec_Conclu}.

\section{\label{sec_model}Development of the 3-D PIC model}
\newpage
\begin{figure}[htbp]
\centerline{\scalebox{1}{\includegraphics{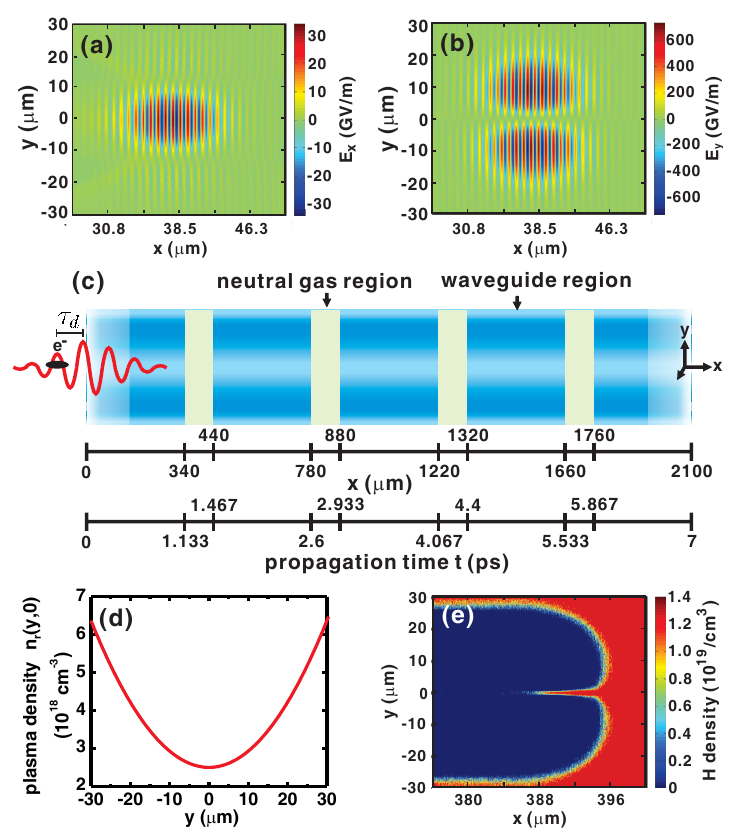}}}
\caption{Snapshots of (a) axial $E_{x}$ and (b) transverse $E_{y}$ electric fields of a 20-fs, 0.5-TW, radially polarized laser pulse with a diameter $w_D=15$ \textmu m; (c) illustration of a density-modulated plasma waveguide, along with the axial position $x$ and the bunch propagation time $t$; (d) transverse plasma density profile $n_{r}(y,0)$ defined for the waveguide regions; (e) ionization of neutral hydrogen gas by the electric field in (a) and (b)} 
\label{Fig_1_DLA_PIC_model}
\end{figure}

The PIC model has been developed using the framework of the commercial software package VORPAL,~\cite{Nieter2004JCP} in which a 3-D Cartesian coordinate system (\textit{x},~\textit{y},~\textit{z}) is defined and the laser pulse propagates along the \textit{x}-axis. As discussed in prior work~\cite{Salamin2006NJF}, the radially polarized electric field in cylindrical coordinates (\textit{r},~$\phi$,~\textit{x}) under paraxial approximation can be expressed in terms of a radial and an axial component, $\textbf{E}=\hat{r} E_{r} + \hat{x} E_{x}$. The radial unit vector $\hat{r}$ in cylindrical coordinates can  be decomposed into the $\hat{y}$ and $\hat{z}$ Cartesian coordinate components:
\begin{equation}\label{unit_vect}
 \hat{r} = \cos \phi~\hat{y} + \sin\phi~\hat{z}
     = \dfrac{y}{(y^{2}+z^{2})^{1/2}}~\hat{y} + \frac{z}{(y^{2}+z^{2})^{1/2}}~\hat{z}.   
\end{equation}   
In our model, the $y$ and $z$ components constituting the radially polarized field are launched into the simulation space as 
\begin{equation}\label{Ey_env}
\begin{split} 
&E_{\alpha} (x,y,z,t)=E_{0}\theta_{0} \dfrac{\alpha}{(y^{2}+z^{2})^{1/2}} \left[ \frac{(y^{2}+z^{2})^{1/2}}{w_{0}} \right] \\ 
& \ \times  \left( \frac{w_{0}}{w(x)} \right)^{2} \exp \left[- \frac{y^{2}+z^{2}}{w(x)^{2}}\right] \mathrm{env}(t) \cos \left( \psi + 2\psi_{G} \right),
\end{split}    
\end{equation}
where $\alpha=y, z$ for the radial field components $E_{y}$ and $E_{z}$, respectively. Here, $w_{0}$ is the focused mode radius and the laser beam size is
\begin{equation}\label{Gau_width} 
   w(x) = w_{0} \sqrt{ 1+\frac{\left(x-x_{f}\right)^{2}}{z_{r}^{2}} }, 
\end{equation}  
which is characterized by the Rayleigh length $z_r= \pi w_0^2/\lambda$ and the focal position $x_{f}$. The beam diffraction angle is $\theta_{0}$~=~$\lambda/ \pi w_0$. The optical phase includes the following contributions:    
\begin{equation}\label{Gau_phase} 
\begin{split} 
& \psi = \psi_{0} + \omega t - k \left(x-x_{f}\right) - \frac{k \left( y^{2}+z^{2}\right) }{2R}, \\
&  R = \left(x-x_{f}\right) + \frac{z_{r}^{2}}{x-x_{f}},  
\end{split}
\end{equation}  
with an absolute phase $\psi_{0}$, $\psi_{G}= \tan^{-1} (x-x_{f})/z_{r}$ commonly referred to as the Gouy phase of a Gaussian beam. Here, $\omega=2 \pi c / \lambda$ is the laser frequency at the wavelength $\lambda$ and $k=2 \pi / \lambda $ is the wave number. The function $\mathrm{env}(t)$ defines a Gaussian laser field envelope as
\begin{equation}\label{Gau_env} 
 \mathrm{env}(t)= \exp \left[-2\,\mathrm{ln}\,2 \frac{\left( t-t_{0}\right)^{2}}{\tau_{p}^{2}} \right],  
\end{equation}  
with a full-width at half-maximum (FWHM) Gaussian pulse duration $\tau_{p}$ and a delay time $t_{0}$. The characteristic field amplitude $E_{0}$ and the laser peak power $P_{0}$ are related by
\begin{equation}\label{E0_P0}
E_{0}= \left(\frac{8c\mu_{0}P_{0}}{\pi w_{0}^{2} \theta_{0}^{2}}\right)^{1/2},
\end{equation}  
in which $\mu_{0}$ is the permeability of free space. The transverse fields $E_{y}$ and $E_{z}$ are defined at the boundary, and the axial field $E_{x}$ and the magnetic field associated with the laser pulse are subsequently calculated from discretized Amp\`ere'€™s and Faraday's laws via the finite-difference time domain (FDTD) method. For a laser pulse with $\lambda=800$ nm that is focused into a FWHM diameter of $w_{D}=15$ \textmu m ($w_{0}=w_{D}/\sqrt{2 \ln 2}\simeq 12.74$ \textmu m) with a duration $\tau_{p}=20$ fs and a peak power of $P_{0}=0.5$ TW, $E_0\simeq 86$\,TV/m gives the maximum radial amplitude $ E_{y,max}=E_{0}\theta_{0}\simeq 737.5$\,GV/m and $E_{x,max}=E_{0}\theta_{0}^{2} \simeq 34.4$\,GV/m.  Figures~\ref{Fig_1_DLA_PIC_model}(a) and (b) show the snapshots of the longitudinal $E_{x}$ and transverse $E_{y}$ components of the electric field of a 20-fs, 0.5-TW, 800-nm radially polarized laser pulse with a diameter $w_{D}=15$ \textmu m in the simulation space. Unless specifically mentioned, those laser pulse parameters are used in the remainder of the simulations described in this work.

For this model, hydrogen is considered to be the gas target for irradiation by the spatially modulated ignitor and spatially uniform heater pulses to produce the density-modulated plasma waveguide, as illustrated in Fig.~\ref{Fig_1_DLA_PIC_model}(c). The simulation starts at the moment when the hydrodynamic expansion of the plasma forms a proper radial plasma density profile $n_{r}(y,z)$ for guiding an injected laser pulse in the longitudinal ($x$) direction. The density profile $n_{r}(y,z)$ of a perfect plasma waveguide that guides a laser pulse in a mode radius $w_{0}$ is defined by a parabolic function:\cite{Serafim2000IEETPS,DurfeeIII1995PRE}
\begin{equation}\label{n_r}
n_{r}(y,z)= n_{p0}+\Delta n \frac{\left( y^{2}+z^{2} \right)}{w_{0}^{2}},
\end{equation}  
where $n_{p0}$ is the on-axis density, $\Delta n=1/ r_{e} \pi w_{0}^{2}$, and $r_{e}$ is the classical electron radius. Figure~\ref{Fig_1_DLA_PIC_model}(d) shows the transverse profile assigned to the waveguide at the beginning of the simulation, with a laser-guided diameter of $w_{D}=15$ \textmu m (or $w_0=12.74$ \textmu m) with $n_{p0}=2.5\times10^{18}$ cm$^{-3}$, while $\Delta n=6.975\times10^{17}$ cm$^{-3}$. In each waveguide section, hydrogen ions and electrons are both defined by the same plasma density function $n_{p}(x,y,z)$. Along the laser propagation direction $x$, $n_{p}(x,y,z)$ for the first waveguide region of length $\lambda_{L,1}$ is a combination of the radial function $n_{r}(y,z)$ with an additional linear density ramp function with a length $L_{r}$, which starts from $x=0$ as 
\begin{equation}\label{n_p}
n_{p}(x,y,z) = \left\{ 
  \begin{array}{l l}
    \left( x / L_{r} \right) n_{r}(y,z) & \ 0 < x \leq L_{r} \\
     n_{r}(y,z)                         & \ L_{r} < x \leq \lambda_{L,1}
  \end{array} . \right.\
\end{equation} 
For a laser beam guided in a waveguide with this density profile, the refractive index $\eta$ is
\begin{equation}\label{eta} 
 \eta (x)=\left( 1- {\omega_{p}(x)}^2 / {\omega}^2 - 8c^2/{\omega}^2{w_0}^2 \right)^{1/2},
\end{equation}
where $\omega_{p}(x)=n_{p}(x,0,0) q_{e}^{2}/\epsilon_{0} m_{e} $ is the on-axis plasma frequency, $\epsilon_{0}$ is the vacuum permittivity, and $m_{e}$ is the electron mass.

In all simulations, the length of density ramp is set to $L_r=150$ \textmu m and the focal position is chosen as $x_f=100$ \textmu m for efficiently coupling the laser pulse into the waveguide.~\cite{Dimitrov2007POP} Next, a neutral gas region of length $\lambda_{H,1}$ is defined with a uniformly distributed hydrogen atom density $n_{a}(x,y,z)=n_{a0}=1.25\times10^{19}$ cm$^{-3}$.  As shown in Fig.~\ref{Fig_1_DLA_PIC_model}(e), hydrogen atoms can be fully ionized by the front foot of a 20-fs, 0.5-TW laser pulse. The results confirm our prediction that hydrogen atoms can be ionized within a few optical cycles and the majority of the pulse experiences a uniformly distributed plasma rather than becoming defocused by ionization-induced refraction.~\cite{IOI_diffract} Since the length of neutral gas regions $\lambda_{H,n}$ is shorter than the Rayleigh length $z_r=637.4$ \textmu m for $\lambda=800$ nm and $w_D=15$ \textmu m, the guiding is sustained over the waveguide length without significant energy leakage. Subsequently, alternating waveguide sections with $n_p(x,y,z)=n_r(y,z)$ and neutral gas sections with $n_a(x,y,z)=n_{a0}$, with lengths $\lambda_{L,n}$ and $\lambda_{H,n}$ respectively for the $n$-th modulation period are defined. The density in the last waveguide region then ramps down to zero at the end of the plasma waveguide.

   The size of the simulation box is $L_x=23.38$ \textmu m in the axial $x$ direction and $L_y \times L_z=60$ \textmu m $\times$ 60 \textmu m in the transverse directions $y$ and $z$. Each simulation has been performed in a moving frame co-propagating with the laser pulse at a speed of light in vacuum $c$. In all simulations, the transverse $y$- and $z$-cell sizes are fixed at $D_y=D_z=400$ nm and $D_x=12.5$ nm (or $D_x=\lambda/64$) for the $x$-cell size. The time step is chosen as $dt=4.16\times10^{-2}$ fs for satisfying the Courant condition. We found the phase velocity $v_{ph}=c/ \eta$ is apparently reduced due to the computational artifact of the FDTD dispersion~\cite{Shlager2003TEEETAP} if a relatively large $D_x$ with respect to the laser wavelength $\lambda$ is assigned in the model. As a result, the DLA dephasing length $L_d=\pi / \vert k - k_{e} \vert$, defined by the laser wave vector $k$ and the electron wave vector $k_e=\omega/v_e$ co-propagating with a velocity $v_e$, is overestimated and leads to computational errors. In addition, a smaller axial cell size $D_x$ assigned in the model results in higher accuracy in calculating the electromagnetic field by the FDTD method. As shown in the previous model~\cite{Lin2012POP}, the theoretical dephasing lengths $L_d$ for the waveguide and neutral gas regions in Fig.~\ref{Fig_1_DLA_PIC_model}(c) are calculated to be $L_d\simeq 340$ \textmu m and $L_d\simeq100$ \textmu m for an electron with a initial kinetic energy $T_0=40$ MeV. Therefore, density modulation lengths $\lambda_{L,n}=340$ \textmu m and $\lambda_{H,n}=100$ \textmu m equal to the corresponding dephasing lengths $L_d$ are assigned to the QPM structure. In addition, the density ramp in the first waveguide region results in a variation of the wave vector $k(x)=2\pi \eta(x)/ \lambda\,\, (0< x \leq L_r)$, such that the section length $\lambda_{L,1}$ has to be carefully adjusted for producing a $\pi$-phase shift. Since the usual Gaussian beam diffraction cannot be neglected, the phase change $\phi$ in the first waveguide region $(0< x \leq \lambda_{L,1}$) can be approximated by
\begin{equation}\label{phi} 
 \phi = \int_0^{\lambda_{L,1}} | k(x) - k_{e} | dx + 2 \tan^{-1} \frac{L_{r}}{z_{r}},
\end{equation}   
where the second term accounts for the Gouy phase change in the density ramp. 
\begin{figure}[tbp]
\centerline{\scalebox{1}{\includegraphics{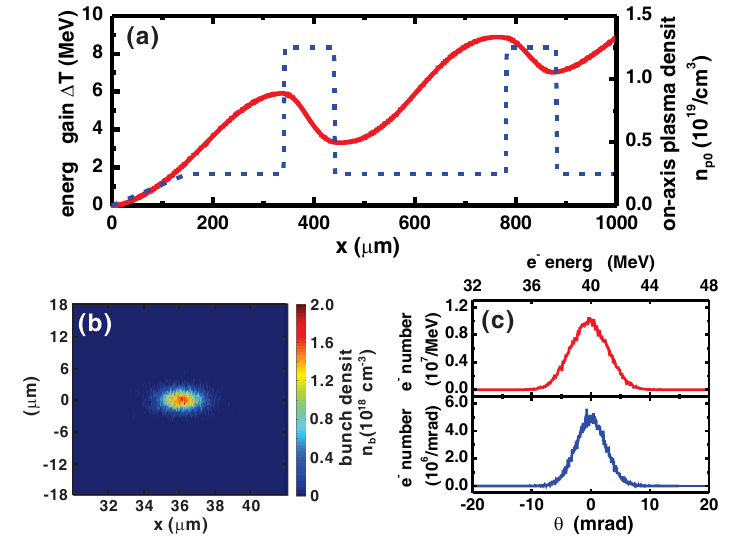}}}
\caption{(a) Energy gain $\Delta T$ and the on-axis plasma density $n_{p0}$ as a function of axial position $x$ for an electron injected at an optimal QPM phase with a initial $T_{0}=40$ MeV. (b) 2-D density distribution of a 6-fs, 40-MeV electron bunch injected in the simulation. (c) The corresponding energy spectrum and the inclination angle $\theta_{y}$ distribution.} 
\label{Fig_2_DLA_PIC_Verify}
\end{figure}
Choosing $n_p(x,0,0)$ in Eq.~(\ref{n_p}) and $\eta(x)$ in Eq.~(\ref{eta}) for the $k(x)$ in Eq.~(\ref{phi}), $\lambda_{L,1}\simeq 340$ \textmu m for $\phi=\pi$. To verify the accuracy of the simulated dephasing length $L_d$ in our model, the dependence of the energy gain $\Delta T$ and the on-axis electron density $n_{p0}(x)$ on axial distance $x$ for an electron with $T_0=40$ MeV is shown in Fig.~\ref{Fig_2_DLA_PIC_Verify}(a). The test electron is injected at a phase $\psi_{i}=5 \pi$ with respect to the axial field $E_{x}\propto \sin \psi$, which fulfills the optimal QPM condition for DLA. Recall $n_{p0}(x)$ for the high density regions, \textit{e.g.} 340 \textmu m $\leq x < 440$ \textmu m, represents the fully ionized plasma density in those original neutral gas regions. The variation of the electron energy gain $\Delta T$ in Fig.~\ref{Fig_2_DLA_PIC_Verify}(a) matches the QPM structure provided by the density modulation, which confirms the FDTD dispersion artifact has been effectively inhibited by setting $D_x=12.5$ nm. Detailed $x$-scale for the density modulation and the laser pulse propagation time $t$ are specified in Fig.~\ref{Fig_1_DLA_PIC_model}(c) for simulations with a 2.1-mm plasma waveguide. 

Fundamental parameters of the electron bunch injected for DLA simulation are determined based on prior LWFA experimental results.~\cite{LWFA_Eng} At LWFA output, the energy spread is approximately 10\% of the average energy, the divergence angle range is 4--10 mrad, and the bunch charge is several pC. The bunch length $L_b$ is predicted to be $\leq \lambda_{p}/4$ (consistent with remaining in the acceleration phase)~\cite{Esarey2009RMP}, and a consistent bunch duration of 2--6 fs ($\tau_b \simeq L_b/c$) has been measured in experiments.~\cite{LWFA_duration} Moreover, the bunch size has been estimated to have 1--2 \textmu m radius and the normalized emittance of the electron bunches has been measured to be between 0.2--2.3 $\pi$-mm-mrad.~\cite{LWFA_emittance} Considering those results from prior LWFA work, a 6-fs, 40-MeV bi-Gaussian electron bunch (having transverse and longitudinal Gaussian shapes) with a bunch diameter $w_b=3$ \textmu m (FWHM), bunch charge of $q_b=5$ pC, and the peak density $n_{b0}=1.6\times10^{18}$ cm$^{-3}$ are defined as default bunch parameters in the model. Figure~\ref{Fig_2_DLA_PIC_Verify}(b) shows the bunch density distribution in the $x$-$y$ plane. The energy spectrum and the distribution of the $y$-dimension inclination angle $\theta_y=P_y/P_x$ (by particle momenta $P_x$ and $P_y$) of the electron bunch are plotted in Fig.~\ref{Fig_2_DLA_PIC_Verify}(c), exhibiting the assigned energy spread of 4 MeV (10\% of initial energy $T_0=40$ MeV) and a divergence angle $\Delta \theta_y$ of 5.9 mrad in FWHM. With the bunch parameters defined above, the default RMS normalized emittance in $y$-dimension is calculated as $\epsilon_{N,y}\simeq 1 \pi$ mm-mrad by the definition:~\cite{Humphries1990Book,Kirby2009PRSTAB}
\begin{equation}\label{Emittance}
   \epsilon_{N,y} = \frac{4}{m_{e}c} 
   \sqrt{ \left\langle y^{2} \right\rangle  \left\langle P_{y}^{2} \right\rangle  -
     \left\langle y P_{y} \right\rangle ^{2}} \: \pi\:\: \mathrm{mm}-\mathrm{mrad} ,   
\end{equation} 
utilizing the particle positions $y$ and momenta $P_y$. With the initial parameters for the default bunch such as $\gamma\simeq 80$, $\sigma_y=w_{b}/(2\sqrt{2 \ln 2})\simeq 1.274$ \textmu m, $\epsilon_{N,y}\simeq \pi$ mm-mrad, and $\beta^{*}=\gamma \sigma_y^2/\epsilon_{N,y}\simeq40.6$ \textmu m, which is relatively short when compared with the 2.1-mm waveguide used in the simulations. The bunch is assigned the same $\theta_z$ divergence distribution and emittance $\epsilon_{N,z}$ in the $z$-dimension as in the $y$-dimension. In data analysis the bunch particles are usually plotted in their $\theta_y-y$ trace space to facilitate the understanding of the change of transverse properties of the bunch in the acceleration process. Because of the azimuthal symmetry of the laser field and the plasma waveguide structure, it is noted that the bunch properties in the $z$-dimension are identical to their counterparts in the $y$-dimension. Therefore, only the bunch transverse properties in $\theta_y-y$ trace space and emittance $\epsilon_{N,y}$ are considered for concise data presentation. Absorbing boundaries for the laser field and all of the particle species are defined around the simulation box. Particles of the injected electron bunch that reach the boundaries are considered to have left the region-of-interest (ROI) of the simulation and are not included in the value of bunch emittance. The ROI of the simulation is equivalent to placing a collimator at the waveguide output in an experiment.

\section{\label{sec_Results} Simulation results and discussion}

\subsection{\label{sec_Delay} Effect of the electron bunch injection delay}

\begin{figure}[tbp]
\centerline{\scalebox{1}{\includegraphics{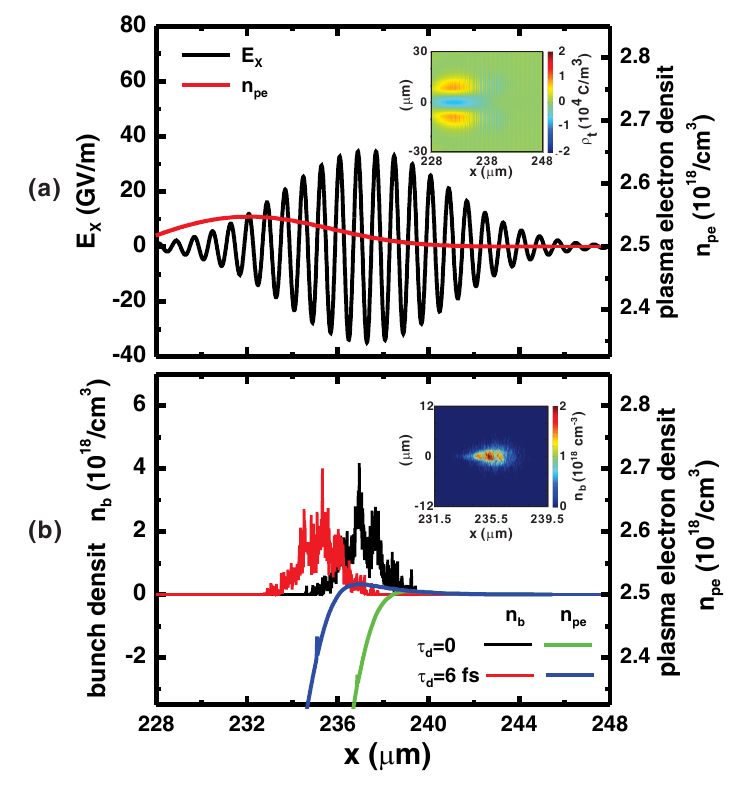}}}
\caption{(a) On-axis axial field $E_{x}$ and the plasma electron density $n_{pe}$ with laser power $P=0.5$ TW. (b) Comparison of the plasma electron density $n_{pe}$ and on-axis bunch density $n_{b}$ when 6-fs bunches, with $\tau_{d}=6.2$ fs and 0, are injected, with other conditions corresponding to (a). Insets in (a) and (b) show the corresponding 2-D total charge $\rho_{t}$ and bunch density distributions. The other parameters are provided in the text.} 
\label{Fig_3_DLA_PlasmaDenst}
\end{figure}

The injection delay $\tau_d$ is defined here as the time delay between the peak of laser pulse envelope and the peak of the electron bunch density distribution. The laser pulse and the electron bunch shape the plasma electron distribution, which in turn produces an electrostatic force that changes the bunch properties. Figure~\ref{Fig_3_DLA_PlasmaDenst}(a) illustrates the variation of the on-axis plasma electron density $n_{pe}(x)$ when a 20-fs, 0.5-TW laser pulse with a beam size $w_D=15$ \textmu m propagates in the first waveguide section illustrated in Fig.~\ref{Fig_1_DLA_PIC_model}(c). 
The corresponding total charge density distribution $\rho_t$ in the $x$-$y$ plane, shown in the inset of Fig.~\ref{Fig_3_DLA_PlasmaDenst}(a), matches the radial field distribution of the laser pulse, in which plasma electrons are depleted in regions of higher $E_{r}$. At this laser intensity, the normalized vector potential is $a_0=q_e E_{r,max}/m_e \omega cm \sim 0.184$. The resulting peak value of the perturbed plasma electron density $n_1(x)=n_{pe}(x)-n_{p0}$ is $n_{1p} \sim 4.6 \times10^{16}$ cm$^{-3}$ (or $n_{1p}/n_{0} \sim 1.8$\%), which produces a radial electrostatic force approximated by $F_s \sim w_b q_e^2 n_1/2 \epsilon_0 \sim 10^{9}$ N. As a result, the electron bunch diverges where the plasma electron density is increased, especially when it is injected near the tailing edge of the laser pulse. Under the same laser and waveguide conditions used in Fig.~\ref{Fig_3_DLA_PlasmaDenst}(a), Fig.~\ref{Fig_3_DLA_PlasmaDenst}(b) shows the variation of $n_{pe}(x)$ when electron bunches with $\tau_d=6.2$ fs and $\tau_d=0$ are injected. The bunch expels the electrons and an ion channel is gradually formed following the front edge of the bunch~\cite{Yoon2012PRSTAB}. At this moment, the ion focusing force at both injection delays increases the peak density of the bunch to $n_{b0}\sim 3 \times10^{18}$ cm$^{-3}$, thus fulfilling the condition for creating an underdense plasma lens. With $\tau_d=6.2$ fs, the bunch density distribution shown in the inset of Fig.~\ref{Fig_3_DLA_PlasmaDenst}(b) has evolved into a ``trumpet'' shape that contains an expanding $head$ region and a $pinch$ region (with a reducing radius), which are typical for an electron bunch propagating in the ion-focusing-regime (IFR).~\cite{Barov1994PRE} However, as shown in Fig.~\ref{Fig_3_DLA_PlasmaDenst}(b), the bunch head experiences a higher on-axis plasma electron density (prepared by the laser pulse) when it is injected with a greater delay $\tau_d$. In this situation, the head erosion of the bunch is amplified by the electrostatic force of the concentrated plasma electrons in addition to the inherent erosion due to a finite emittance.~\cite{Buchanan1987PF}       

The electron energy gain varies with the injection delay $\tau_d$ because of the walk-off effect between the laser pulse and the electron bunch. The energy gain in QPM DLA can be estimated by~\cite{Lin2012POP}  
\begin{equation}\label{delta_T}
 \Delta T \simeq C_{e} C_{qpm} C_{\mathrm{env}} \, q_{e}E_{x,max}L_{wg}, 
\end{equation} 
where $C_e=\int_0^{\pi} \sin (\phi) d\phi/\pi \simeq 0.637$ is the half-cycle average of the electric field amplitude within one dephasing length, $C_{qpm}$ is the correction that accounts for the QPM process:   
\begin{equation}\label{C_qpm}
 C_{qpm}= (\lambda_{L,n}-\lambda_{H,n})/(\lambda_{L,n}+\lambda_{H,n}),
\end{equation}   
and the average amplitude of the effective field envelope is    
\begin{equation}\label{C_env}
 C_{\mathrm{env}}=\dfrac{1}{L_{wg}}\int_ {x_{\mathrm{int}}}^{x_{\mathrm{fin}}} \exp \left[ \, - 2 \ln 2 ( \frac{x}{L_{en}} )^2 \right] \,d x\\,
\end{equation}     
which transforms the envelope function in Eq.~(\ref{Gau_env}) into a spatially variable function with an effective width $L_{en}$ related to the laser pulse length $L_p=c \tau_p$ by       
\begin{equation}\label{L_en}
 L_{en}= \frac{v_{e}}{v_{e}- v_{g,avg}} L_{p}.
\end{equation} 
The average laser pulse group velocity $v_{g,avg}=(\lambda_{L,n}v_{g,L}+\lambda_{H,n}v_{g,H})/(\lambda_{L,n}+\lambda_{H,n})$ is calculated with the group velocities $v_{g}(x)=c \eta (x)$ in low- and high-density regions, denoted by  $v_{g,L}$ and $v_{g,H}$, respectively. The initial electron position $x_{\mathrm{ini}}$ with respect to the effective field envelope is determined by the injection delay $\tau_d$:       
\begin{equation}\label{x_ini}
 x_{\mathrm{ini}}= - \tau_d c \frac{v_{e}}{v_{e}- v_{g,avg}},
\end{equation} 
and the final electron position is $x_{\mathrm{fin}}=x_{\mathrm{ini}}+L_{wg}$. An electron which experiences a symmetrically distributed acceleration field envelope with along the waveguide axis acquires the highest energy $\Delta T_{max}$. According to Eq.~(\ref{L_en}), the delay $\tau_d'$ associated with $\Delta T_{max}$ can be estimated by
\begin{equation}\label{tou_max} 
\tau_{d}' \simeq \frac{v_{e}- v_{g,avg}}{v_{e}}\frac{L_{wg}}{2c}.
\end{equation}
For the waveguide shown in Fig.~\ref{Fig_1_DLA_PIC_model}(c) with $L_{wg}=2.1$ mm, $v_{g,avg}\simeq 0.9982\,c$, and $v_e \simeq c$ gives $\tau_d' \simeq 6.2$ fs and, consequently, $x_{\mathrm{int}}\simeq -1.05$ mm, $x_{\mathrm{fin}}\simeq1.05$ mm, and $L_{en}\simeq 3.39$ mm (at $\tau_{p}=20$ fs) lead to $C_{\mathrm{env}}\simeq 0.9574$. The correction factors is $C_{qpm} \simeq 0.5455$ for the first modulation period and should be modified to $C_{qpm}=1$ for the last waveguide region. As a result, $\Delta T_{max}\simeq 27.2$ MeV is estimated by Eq.~(\ref{delta_T}) for an electron injected at $\tau_d'=6.2$ fs. This injection delay can be chosen such that a large fraction of the electrons around the density peak region can experience efficient DLA. With a reduced injection delay of $\tau_d'=0$, $x_{\mathrm{int}}\simeq 0$ and $x_{\mathrm{fin}}\simeq 2.1$ mm yield the lowered $C_{\mathrm{env}}\simeq 0.8478$ and $\Delta T_{max}\simeq 24.1$ MeV. Therefore, the maximum energy of the accelerated electrons is reduced when the bunch is injected with a shorter delay $\tau_d$.   

\begin{figure}[tbp]
\centerline{\scalebox{1}{\includegraphics{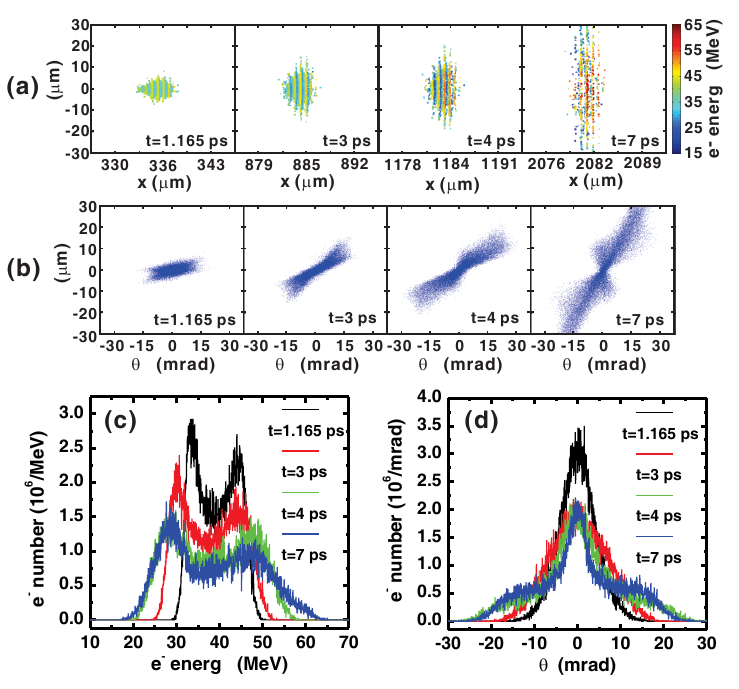}}}
\caption{Variation of the (a) electron distribution ($|z|\leq0.4$ \textmu m), (b) trace space, (c) energy spectrum, and (d) $\theta_{y}$ distribution for an bunch injected with $\tau_{d}=6.2$fs, $T_{0}=40$ MeV, $\tau_{b}=6$ fs, and propagates in a 2.1-mm long plasma waveguide.} 
\label{Fig_4_DLA_Delay6}
\end{figure}

\begin{figure}[tbp]
\centerline{\scalebox{1}{\includegraphics{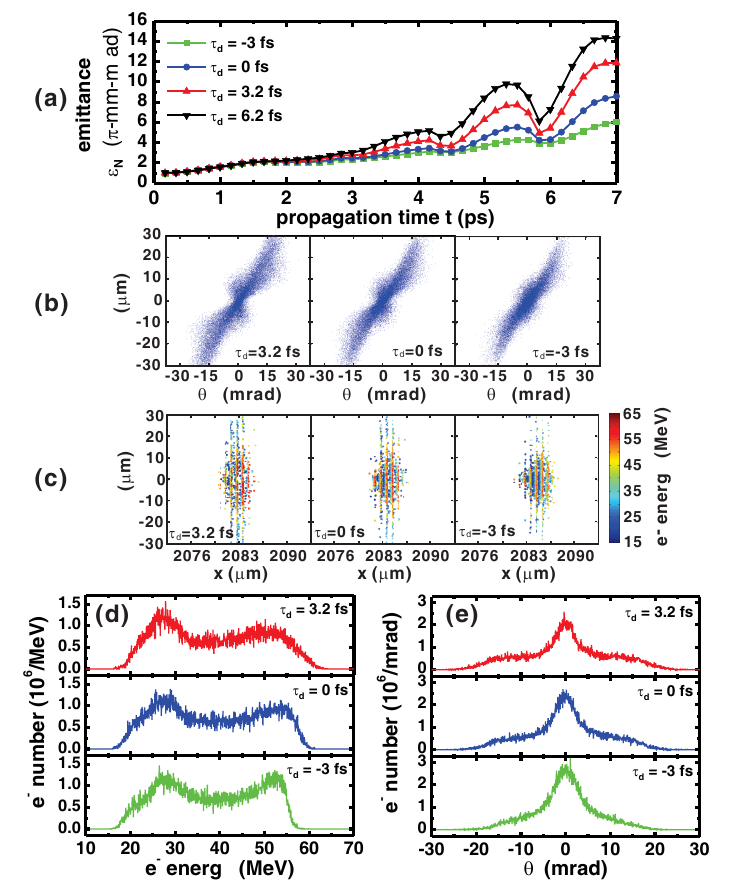}}}
\caption{(a) Bunch emittance $\epsilon_{N,y}$ as a function of the propagation time $t$ for different time delays $\tau_{d}=6.2$ fs, 3.2 fs, 0, and -3 fs. Comparison for the final (b) trace space distributions, (c) electron distributions, (d) energy spectra, and (e) $\theta_{y}$ distributions for bunches injected with $\tau_{d}=3.2$ fs, 0, and -3 fs.} 
\label{Fig_5_DLA_DelayEmit}
\end{figure}

We next discuss the bunch characteristics following DLA when $\tau_d=6.2$ fs, with the other default bunch parameters introduced previously. The spatial particle distribution (within $|z|\leq 0.4$ \textmu m),  $\theta_{y}-y$ trace space, energy spectrum, and $\theta_{y}$ distribution for the bunch electrons are shown in Fig.~\ref{Fig_4_DLA_Delay6}. The series of snapshots in time shown in Fig.~\ref{Fig_4_DLA_Delay6}(a) illustrates the effect of the electron injection phase on the acceleration or deceleration process, resulting the gradual broadening of the energy spectrum, shown in Fig.\ref{Fig_4_DLA_Delay6}(c). The radial Lorentz force $F_{r}\propto q_{e}E_{r}$ also focuses or defocuses the bunch electrons according to their injection phases with respect to the radial field $E_{r}$.\cite{Lin2012POP} However, the radial electrostatic force resulting from the concentrated on-axis plasma electrons shown in Fig.~\ref{Fig_3_DLA_PlasmaDenst}(b) acts to increase the divergence of the electron bunch along the entire propagation distance in the waveguide. Since the axial and radial field are out of phase by $\pi/2$, off-axis electrons injected near the optimal axial acceleration phases predominantly remain in the defocusing phases in the QPM process. Many of those electrons thus move to the outer radial region and, as a result, the electron number at the high-energy end of the spectrum in Fig.~\ref{Fig_4_DLA_Delay6}(c) is significantly reduced. The maximum energy gain is $\Delta T_{max}\simeq 25$ MeV, which is close to the previously predicted value. In contrast, off-axis electrons injected near the axial deceleration phase are primarily focused by the radial Lorentz force. This focusing force helps to confine those electrons axially, so that they are more effectively decelerated by the axial field along the propagation. 
As a result, the final spectrum in Fig.\ref{Fig_4_DLA_Delay6}(c) shows a higher number of electrons at low energies and becomes asymmetric with respect to the initial energy $T _{0}=40$ MeV. The bunch electrons stay within the radial position $r\leq 9$ \textmu m (the peak of the radial field when $w=15$ \textmu m), experiencing focusing from the laser ponderomotive force. The confinement effect becomes prominent at a distance $x\sim 1200$ \textmu m (or at the propagation time $t=4$ ps) when the electron bunch is synchronized with the laser pulse. Consequently, bunch electrons within $r\leq 9$ \textmu m are better collimated. As shown in the $\theta_{y}$-$y$ trace space in Fig.~\ref{Fig_4_DLA_Delay6}(b), particles in the vicinity of the on-axis region ($y=0$) exhibit smaller values of $\theta_{y}$, especially when $t=7$ ps. This property can also be observed from the $\theta_{y}$ distribution shown in Fig.~\ref{Fig_4_DLA_Delay6} (d), in which the collimation effect provided by the laser ponderomotive force is evident. However, the bunch still has an overall tendency to diverge, and its emittance $\epsilon_{N,y}$ increases from $1 \pi$-mm-mrad at the point of injection to approximately $14 \pi$-mm-mrad at the output ($x=2.1$ mm), as shown in Fig.~\ref{Fig_5_DLA_DelayEmit}. The periodic change in $\epsilon_{N,y}$ results from the contribution by the electrons injected in the defocusing phase. The electrons follow a periodic trajectory in the radial direction due to the periodic phase change of the radial force direction. As those electrons gradually move to the outer radial region and are driven by stronger radial force, $\epsilon_{N,y}$ changes more rapidly and oscillates with a greater amplitude with the increased propagation time $t$.      

Reducing the injection delay $\tau_{d}$ helps to mitigate the bunch divergence. As shown in Fig.~\ref{Fig_3_DLA_PlasmaDenst}(a), the perturbed on-axis plasma density $n_{1}(x)$ is reduced near the leading edge of the laser pulse. Electron bunches injected with a smaller $\tau_d$ experience a reduced defocusing force from the perturbed background plasma. The ponderomotive force of the laser also peaks at $\tau_d=0$; therefore, the confinement force increases with a smaller injection delay. To improve the emittance and collimation after DLA, selected injection delays $\tau_d=3.2$ fs, $0$, and $-3.2$ fs are assigned to the bunches, with the remaining bunch parameters the same as in the previous analysis. As shown in Fig.~\ref{Fig_5_DLA_DelayEmit}(a), the final emittance $\epsilon_{N,y}$ and the amplitude of its temporal oscillation are reduced at smaller injection delays $\tau_d$. Because of the walk-off effect, the electron bunch overtakes the laser pulse and experiences the decreasing radial field at the leading edge of laser pulse as it approaches the waveguide output. With a smaller $\tau_d$, the effective radial Lorentz force $F_{r}$ experienced by the electrons is further decreased, which explains the reduced oscillation of $\epsilon_{N,y}$. This condition corresponds to the final $\theta_{y}$-$y$ trace space distributions shown in Fig.~\ref{Fig_5_DLA_DelayEmit}(b), in which particles lying around $y=\pm 10$ \textmu m are less scattered when $\tau_d$ is smaller. The improved collimation of the bunch at a smaller delay $\tau_{d}$ can also be observed from the comparison of the particle distributions in Fig.~\ref{Fig_5_DLA_DelayEmit}(c) and the $\theta_{y}$ distribution in Fig.~\ref{Fig_5_DLA_DelayEmit}(e) at three different injection delays. As more electrons remain in the region where the laser field is intense, the fraction of electrons accelerated to higher energies is increased. The comparison of electron energy spectra shown in Fig.~\ref{Fig_5_DLA_DelayEmit}(d) indicates an increased electron number in the range $50-60$ MeV with a reduced $\tau_d$ that is attributed to the reduced bunch divergence. However, the maximum energy in the spectrum drops from 65 MeV when $\tau_d=6.2$ fs to 55 MeV when $\tau_d=-3.2$ fs, which can be attributed to a reduced $C_{\mathrm{env}}$ correction factor for electrons injected with a smaller delay $\tau_d$. The results presented in this section with a short bunch duration $\tau_b$ show that the radial Lorentz force predominately drives the change of bunch emittance $\epsilon_{N,y}$ in DLA. As more electrons are defocused to the regions with stronger radial field, the growth of emittance $\epsilon_{N,y}$ is enhanced accordingly. The collimation of the bunch after DLA can be improved by selecting a smaller injection delay $\tau_d$. 
     
\subsection{\label{sec_Length} Effect of the electron bunch length} 

\begin{figure}[tbp]
\centerline{\scalebox{1}{\includegraphics{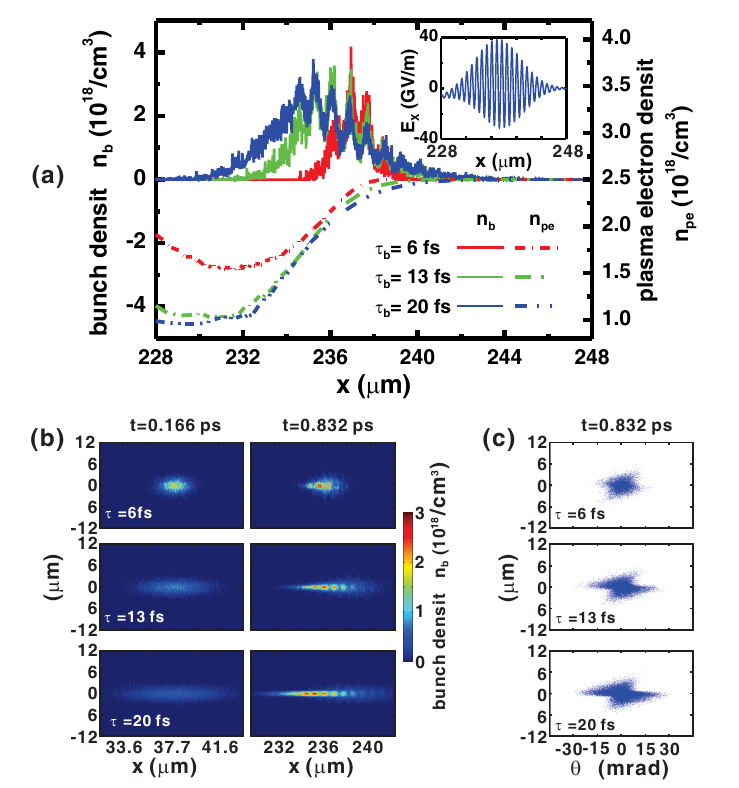}}}
\caption{(a) Comparison of the on-axis bunch density $n_{b}$ and plasma electron density $n_{pe}$ at $t=0.832$ ps for bunches with durations $\tau_{b}=6$ fs, 13 fs and 20 fs are injected at $\tau_{d}=0$. The corresponding (b) variation of the 2-D bunch density from $t=0.166$ ps to $t=0.832$ ps and (c) the trace space distributions at $t=0.832$ ps. The other parameters are given in the text.} 
\label{Fig_6_DLA_EbmLen}
\end{figure}

Results described in the previous section indicate a trend of increasing divergence in DLA of short electron bunches. With a fixed injection delay $\tau_d=0$ and bunch charge of $q_b=5$ pC, Fig.~\ref{Fig_6_DLA_EbmLen}(a) shows the comparison of on-axis plasma density $n_{pe}(x)$ when bunch duration is set to $\tau_{b}=6$ fs, 13 fs and 20 fs, while the rest of the bunch and laser parameters are kept the same as in the previous analysis. Regardless of the bunch duration $\tau_b$, the reduction of the plasma electron density $n_{pe}(x)$ is always initiated at the leading edge of the bunch. For the 6-fs electron bunch having a length $L_b=\tau_b c=1.8$ \textmu m, the majority of bunch electrons do not experience a strong focusing force from the created ion channel since the variation of $n_{pe}(x)$ is of order $\lambda_p/4=\pi c/2 \omega_{p0}\simeq 5.3$ \textmu m near the pulse falling edge. Therefore, for increased durations of the electron bunch of $\tau_{b}=13$ fs and 20 fs, the corresponding bunch lengths $L_b=3.9$ \textmu m and 6 \textmu m are closer to the value of $\lambda_p/4$, such that more bunch electrons can be confined in the created ion channel. Consequently, the collimation and emittance of the DLA-accelerated bunch can be improved. The ion-focusing effect also rapidly increases the density of the injected bunch when $\tau_b=13$ fs and 20 fs, as shown in Fig.~\ref{Fig_6_DLA_EbmLen}(b). 
For example, the peak density $n_{b0}=4.8 \times 10^{17}$ cm$^{-3}$ for the 20-fs bunch is increased to about $n_{b0}=3 \times 10^{18}$ cm$^{-3}$ after propagating for 237 \textmu m in the plasma and the main $body$ region~\cite{Barov1994PRE} with a constant radius occurs at the tailing edge of the bunch. The increased density for bunches with $\tau_b=13$ fs and 20 fs also enhances the ion-focusing force, which can be understood from the further reduced $n_{pe}(x)$ in Fig.~\ref{Fig_6_DLA_EbmLen}(a). Comparing the trace space results in Fig.~\ref{Fig_6_DLA_EbmLen}(c), larger $\theta_{y}$ values are characteristic for the electrons with a larger $\tau_b$, since they experience an increased ion-focusing force in the trailing edge of the bunch. The trace spaces for $\tau_b=13$ fs and 20 fs are those typical for the IFR region~\cite{Su1990PRA}, in which many particles at $y>0$ positions are associated with $\theta_{y}<0$ (and vice versa), indicating a strong focusing force on the bunch.          
  
\begin{figure}[tbp]
\centerline{\scalebox{1}{\includegraphics{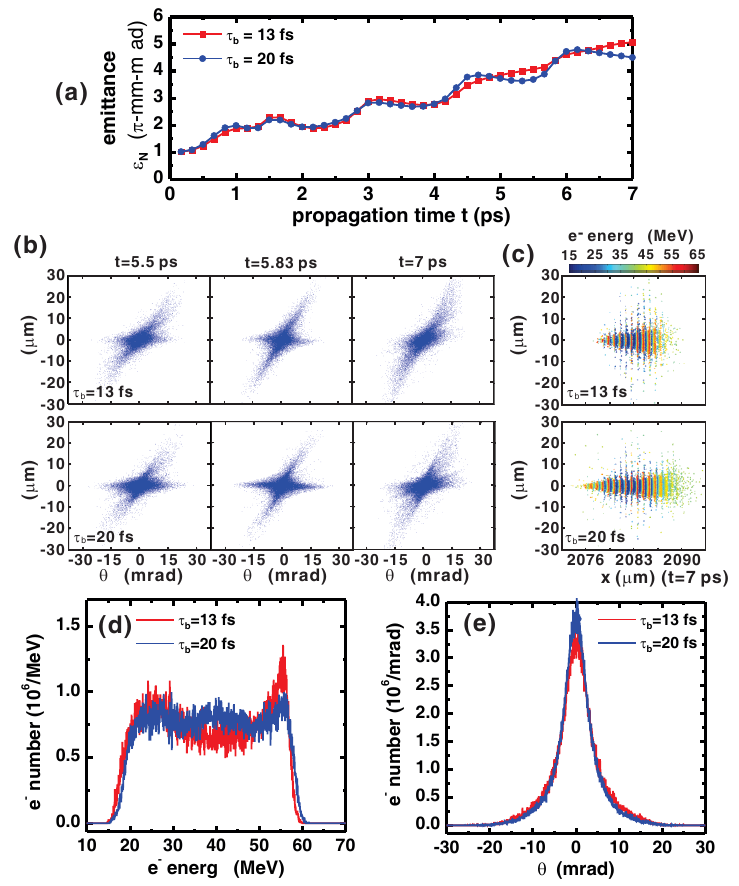}}}
\caption{(a) Bunch emittance $\epsilon_{N,y}$ as a function of propagation time $t$ for bunches with durations $\tau_{b}=13$ fs and 20 fs. (b) Sampled trace space distributions and final (c) electron distributions, (d) energy spectra and (e) $\theta_{y}$ distributions.} 
\label{Fig_7_DLA_EbmLenEmit}
\end{figure}

Comparing Figs.~\ref{Fig_5_DLA_DelayEmit}(a) with \ref{Fig_7_DLA_EbmLenEmit}(a), the emittance $\epsilon_{N,y}$ can be considerably reduced by increasing the bunch duration to $\tau_b=13$ fs and 20 fs with the same delay time $\tau_d=0$. The bunch electrons can be more concentrated at the waveguide center, as shown in Fig.\ref{Fig_7_DLA_EbmLenEmit}(b) at those longer bunch durations, which is attributed to the enhanced ion-focusing effect. In contrast to the trace space results obtained for short bunches, electrons concentrated in $|y|\leq$ 5\textmu m in Fig.~\ref{Fig_7_DLA_EbmLenEmit}(b) can have a large value of $\theta_{y}$ due to the increased  transverse momentum $P_{y}$ driven by the ion focusing force. 
Examining the trace spaces at $t=5.5$ ps and $t=5.83$ ps, the range of $\theta_{y}$ is increased when the bunch  propagates in the high-density regions, where the ion focusing force is enhanced by a higher plasma density $n_{p}$. Many of the bunch electrons are collimated in the next low-density region, as evidenced by the reduced range of $\theta_{y}$ between $t=5.83$ ps and $t=7$ ps, shown in Fig.~\ref{Fig_7_DLA_EbmLenEmit}(b). Consequently, the bunch emittance $\epsilon_{N,y}$ in Fig.~\ref{Fig_7_DLA_EbmLenEmit}(a) increases during propagation in high-density regions and decreases in low-density regions, especially when $\tau_b=20$ fs. The increasing emittance $\epsilon_{N,y}$ in the first low-density (waveguide) region is an exception, in which $\theta_{y}$ (or $P_{y}$) continuously increases, as illustrated in Fig.~\ref{Fig_6_DLA_EbmLen}(c) for $t=0.832$ ps. As the bunch duration $\tau_b$ increases, electrons experience a large fraction of the varying-strength laser field over the entire acceleration length. The final bunch particle distributions for $\tau_b=13$ fs and 20 fs in Fig.~\ref{Fig_7_DLA_EbmLenEmit}(c) show that more electrons at the leading and trailing edges cannot be effectively accelerated/decelerated when the bunch duration becomes comparable to the laser pulse duration of $\tau_p=20$ fs. Therefore, the final energy spectra in Fig.~\ref{Fig_7_DLA_EbmLenEmit}(d) become more uniform with increased bunch duration $\tau_b$. The maximum energy gain, however, drops to approximately $\Delta T_{max}\simeq 20$ MeV due to the biased axial field amplitude illustrated in the inset of Fig.~\ref{Fig_6_DLA_EbmLen}(a), in which a 5-GV/m field resulting from the density variation $n_{p0}(x)$ is superimposed on the axial field $E_{x}$. This net positive electric field within the phase region $0-\pi/2$ of the excited plasma wave corresponds to the axial decelerating field in a plasma wakefield accelerator,~\cite{Ruth1985PA} which can also drive the Ohmic dissipation~\cite{Buchanan1987PF} of the electron energy. When an additional DLA field is present, the local wakefield cancels a part of the DLA gradient, such that the maximum energy gain  $\Delta T_{max}$ is consequently reduced. The narrow $\theta_{y}$ distributions shown in Fig.~\ref{Fig_7_DLA_EbmLenEmit}(e) are consistent with the improved emittance $\epsilon_{N,y}$ when the bunch duration is increased.                

\begin{figure}[tbp]
\centerline{\scalebox{1}{\includegraphics{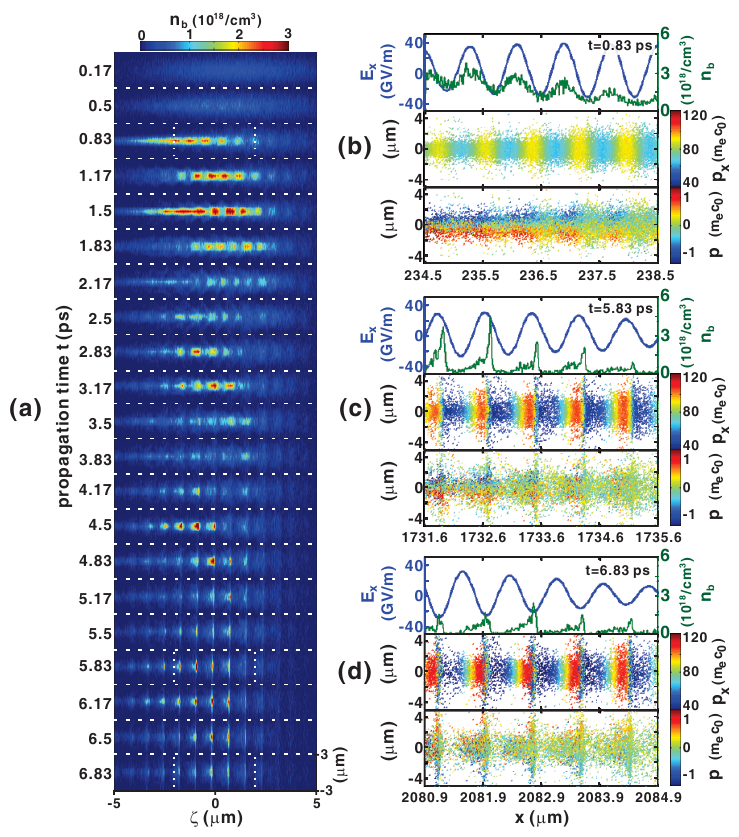}}}
\caption{(a) Sampled 2-D bunch density variation in the entire propagation for a 20-fs injected bunch. The corresponding on-axis axial field $E_{x}$, bunch density $n_{b}$, axial $P_{x}$ and transverse $P_{y}$ momentum distributions at (b) $t=0.83$ ps, (c) $t=5.83$ ps, and (d) $t=6.83$ ps.} 
\label{Fig_8_DLA_EbmLenDenst}
\end{figure}

The formation of density peaks in QPM of DLA~\cite{Yoon2012PRSTAB,DLA_OSWS}, or microbunches, becomes prominent when a long bunch is injected. Since most of the bunch electrons can be confined in the ion channel over a long distance, a sufficient time exists during the DLA process for this density modulation to be realized. In a moving coordinate of the simulation box $\zeta=x-ct$, Fig.~\ref{Fig_8_DLA_EbmLenDenst}(a) shows the evolution of the bunch density throughout its propagation in the 2.1-mm long waveguide. In the early phase of propagation, the density modulation results from the focusing and defocusing of the bunch by the radial Lorentz force $F_{r}$. The on-axis bunch density $n_b$ in the central axial region and the electron momenta ($P_{x}$ and $P_{y}$) at $t=0.83$ ps are shown in Fig.~\ref{Fig_8_DLA_EbmLenDenst}(b). This radial force induces a periodic change of the electron transverse momentum $P_y$, and the bunch density $n_b$ peaks at the phases where electrons are focused (the corresponding regions with $P_{y}>0$ are shown in red and and with $P_{y}<0$ are shown in blue). 
As the electrons are continuously accelerated/decelerated in the DLA process, the increased axial velocity difference between the electrons then gradually starts to dominate, similar to the effect seen in the traveling wave tubes.~\cite{DLA_OSWS,Tsimring2007Book} From Fig.~\ref{Fig_8_DLA_EbmLenDenst}(c), at $t=5.83$ ps the bunching happens at the regions where the acceleration phase (red) switches to the retarding phase (blue) with a period of 800 nm, equal to the laser wavelength. The peak density of the microbunches can be approximately one order of magnitude higher than the original peak density $n_{b0}=4.8 \times 10^{17}$ cm$^{-3}$. The density of the microbunches continues to change as they propagate. At $t=6.83$ ps, the densities of the microbunches drop, mainly driven by the defocusing of electrons by the radial Lorentz force $F_r$. Therefore, the density modulation in DLA is a highly nonlinear process that results from the combined effect of the radial force and the axial momentum (or velocity) modulation on the bunch. Comparing Figs.~\ref{Fig_8_DLA_EbmLenDenst}(b) with (c), it can be observed that the bunch density peaks in the phases offset by $\pi$ with respective to the axial electron momentum ($P_{x}$) modulation. Therefore, the phase of the $P_{x}$ modulation can be used as a signature that identifies the dominant bunching mechanism in DLA.  
 
\subsection{\label{sec_Size} Effect of the transverse electron bunch size}
  
The finite diameter of the laser beam limits the size of the effective radial region and the efficiency of DLA because of the reduced axial field available to the off-axis electrons.~\cite{DLA_OSWS,Lin2006POP} On the other hand, the density modulation is enhanced as the off-axis electrons experience a stronger radial focusing/defocusing Lorentz force $F_{r}\propto q_{e}E_{r}$. To understand the effect of the electron bunch diameter on DLA, bunches with two diameters ($w_b=9$ \textmu m and 15 \textmu m), fixed duration $\tau_b=6$ fs and total charge $q_b=5$ pC are injected. In both cases the bunches are assigned the same initial emittance $\epsilon_{N,y}\simeq 1 \pi$-mm-mrad, so that the divergence angles are $\Delta \theta_{y}\simeq 1.96$ mrad and 1.177 mrad for $w_b=9$ \textmu m and 15 \textmu m, respectively. 

\begin{figure}[tbp]
\centerline{\scalebox{1}{\includegraphics{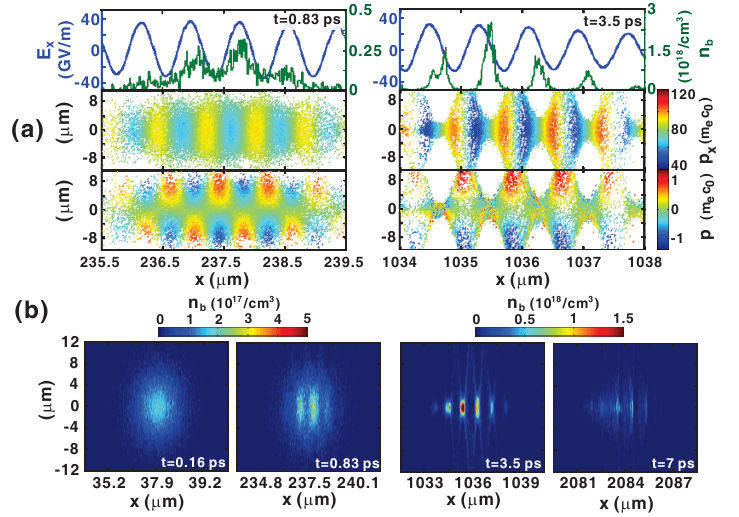}}}
\caption{(a) Comparison of on-axis axial field $E_{x}$, bunch density $n_{b}$, axial $P_{x}$ and transverse $P_{y}$ momentum distributions at $t=0.83$ ps and 3.5 ps for a bunch of size $w_{b}=9$ \textmu m. (b) Sampled 2-D bunch density variation in the entire propagation. The other parameters are provided in the text.} 
\label{Fig_9_DLA_EbmSize}
\end{figure}
In Fig.~\ref{Fig_9_DLA_EbmSize}(a), a comparison is provided for the change of on-axis bunch density $n_b$ and electron momenta at $t=0.83$ ps and $t=3.5$ ps for an injected bunch with $w_b=9$ \textmu m. The radial force $F_r$ induces a periodic change of the electron transverse momentum $P_y$ at $t=0.83$ ps, from which the primary focusing and defocusing phases are determined. The focused electrons subsequently become concentrated in the center of the waveguide and experience a reduced radial force. At $t=3.5$ ps, densities of the microbunches reach their peak values, up to one order of magnitude higher that the injected peak density of $n_{b0}=1.78 \times 10^{17}$ cm$^{-3}$. The diameter of the microbunches is reduced to approximately 2 \textmu m as the electrostatic forces is balanced by the radial force~\cite{DLA_OSWS}. By comparing the plots of $n_b$ and the electron axial momentum $P_x$, it can be understood that the radial bunching phases coincide with the axial defocusing phases. As a consequence, it can be observed that densities of the microbunches drops after $t=3.5$ ps due to the de-bunching effect arising from the $P_x$-modulation. Figure~\ref{Fig_9_DLA_EbmSize}(b) summarizes the variation of the bunch density during the entire propagation through the waveguide. At $t=7$ ps, when the bunch arrives the exit of the waveguide, the scattered axial electron distribution is consistent with the de-bunching effect that reduces the microbunch density. By changing the waveguide length to 1.05 mm, microbunches with the highest available peak densities can be produced. Electrons within each microbunch exhibit a broad energy spectrum, however. For example, the energy spread is approximately 25 MeV for a microbunch at $t=3.5$ ps in Fig.~\ref{Fig_9_DLA_EbmSize}(b). 

For large bunch transverse size, the variation of emittance $\epsilon_{N,y}$ is directly related to the effect of focusing and defocusing of the bunch by the radial force $F_{r}$.
\begin{figure}[tbp]
\centerline{\scalebox{1}{\includegraphics{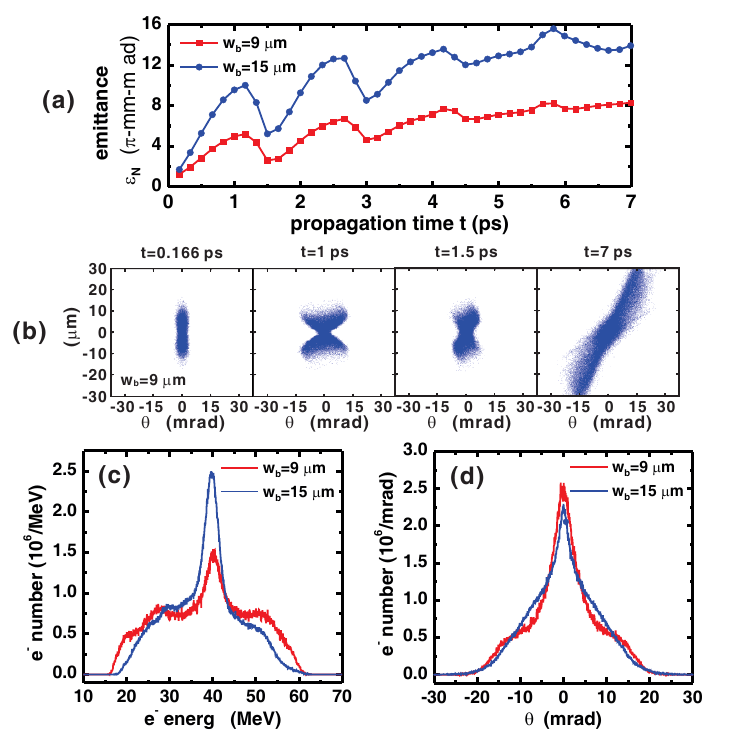}}}
\caption{(a) Bunch emittance $\epsilon_{N,y}$ as a function of propagation time $t$ for bunches with sizes $w_{b}=9$ \textmu m and 15 \textmu m. (b) Sampled trace space distributions for $w_{b}=9$ \textmu m. Final (d) energy spectra and (e) $\theta_{y}$ distributions for $w_{b}=9$ \textmu m and 15 \textmu m.} 
\label{Fig_10_DLA_EbmSizeEmit}
\end{figure}
Figure~\ref{Fig_10_DLA_EbmSizeEmit}(a) shows the change of emittance $\epsilon_{N,y}$ when bunches of $w_b=9$ \textmu m and 15 \textmu m are injected. In both cases, the emittance $\epsilon_{N,y}$ changes periodically and tends to increase as the bunch propagates through the waveguide. The periodic focusing and defocusing of many of the electrons by the radial force, illustrated in Fig.~\ref{Fig_9_DLA_EbmSize}(a), accounts for the rapid change of $\epsilon_{N,y}$ before $t=3.5$ ps. In this situation, the numbers of electrons having $P_y>0$ and $P_y<0$ are similar. When $w_b=9$ \textmu m, the trace spaces between $t=0.166$ ps and 1.5 ps in Fig.~\ref{Fig_10_DLA_EbmSizeEmit}(b) become symmetrically distributed with respect to the two axes. The increasing magnitude of $|P_y|$ of those electrons, while being accelerated/decelerated by the radial force, leads to a broadened $\theta_{y}$ distribution at $t=1$ ps. When the bunch propagates into the next high-density region, the reversal of the radial force $F_{r}$, due to the change of $E_{r}$ pointing, decreases $|P_y|$ and the populated range of $\theta_{y}$. At $t=3.5$ ps, the variation of $\epsilon_{N,y}$ is determined mostly by the electrons in the defocusing phases of the radial force, as illustrated in Fig.~\ref{Fig_9_DLA_EbmSize}(a). The oscillations of $\epsilon_{N,y}$ are suppressed after $t=3.5$ ps as more of the defocused electrons have left the region of the high electric field of the laser pulse and, finally, results in a broadly distributed trace space at $t=7$ ps, as shown in Fig.~\ref{Fig_10_DLA_EbmSizeEmit}(b). 

When $w_b=15$ \textmu m, the enhanced effect of the radial force $F_{r}$ leads to the further increased $\epsilon_{N,y}$ along the propagation. The DLA efficiency is reduced, as shown by the final energy spectra in Fig.~\ref{Fig_10_DLA_EbmSizeEmit}(c), with $w_b=9$ \textmu m and 15 \textmu m, where a large fraction of the bunch electrons still have energies around 40 MeV. Increased bunch sizes also lead to greater divergence angles $\Delta \theta_{y}$, as shown in the final $\theta_{y}$ distributions in Fig.~\ref{Fig_10_DLA_EbmSizeEmit}(c). The results presented in this section show that increasing the injected bunch transverse size negatively impacts DLA performance. However, the formation of microbunches can be enhanced by choosing a larger bunch diameter, whereby the electrons experience a greater radial force that drive the density modulations.   
  
\subsection{\label{sec_PWRLWD} Effect of the laser power and waveguide length}              
\begin{figure}[tbp]
\centerline{\scalebox{1}{\includegraphics{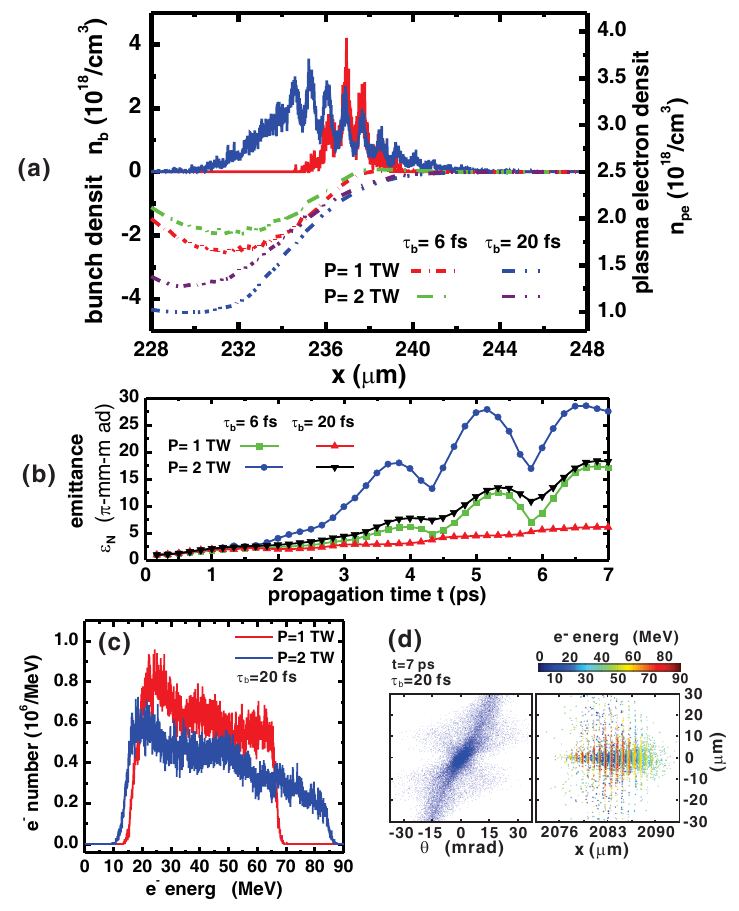}}}
\caption{Effect of the laser power ($P=1$ TW and 2 TW). (a) Comparison of the on-axis bunch density $n_{b}$ and plasma electron density $n_{pe}$ at $t=0.832$ ps for bunches injected at $\tau_{d}=0$ with durations $\tau_{b}=6$ fs and 20 fs. (b) The emittance $\epsilon_{N,y}$ as a function of propagation time $t$. (c) Final energy spectra for bunch with $\tau_{b}=20$ fs. (e) The final trace space and electron distributions for bunch with $\tau_{b}=20$ fs with laser power $P=2$ TW. The waveguide length $L_{wg}=2.1$ mm is fixed for all cases.} 
\label{Fig_11_DLA_EbmPwr}
\end{figure}

The energy gain scales as $\Delta T \propto q_{e}E_{x,max}L_{wg}$ for DLA in a plasma waveguide. By increasing the laser power $P \propto E_{x,max}^2$ or the waveguide length $L_{wg}$, a higher energy gain can be achieved through DLA. To understand how the bunch properties change with the laser power, two laser pulse powers ($P=1$ TW and 2 TW) are used to accelerate the electron bunches with two different durations ($\tau_b=6$ fs and 20 fs), injected with a fixed delay $\tau_d=0$ into waveguides with length $L_{wg}=2.1$ mm. In Fig.~\ref{Fig_11_DLA_EbmPwr}(a) the variation of the on-axis plasma electron density $n_{pe}(x)$ when the bunches propagate in the first waveguide section is shown. The plasma density perturbation $n_1(x)$ can be inhibited by increased the laser pulse power, since a stronger laser ponderomotive force overcomes the electrostatic force from the bunch that acts to expel the plasma. The ion focusing force acting on the electrons is lowered accordingly. The increased radial field at higher laser power also leads to a greater defocusing of a fraction of the bunch electrons. As a result, the emittance $\epsilon_{N,y}$ increases with increased laser power, as shown in Fig.~\ref{Fig_11_DLA_EbmPwr}(b). When $\tau_b=6$ fs and $P=2$ TW, the large final emittance $\epsilon_{N,y}\simeq27.5 \pi$-mm-mrad leads to a 40\% loss of bunch electrons in the ROI. Therefore, only bunches of $\tau_b=20$ fs still retain an acceptably low final emittance when the laser power is increased up to $P=2$ TW. Fig.~\ref{Fig_11_DLA_EbmPwr}(c) shows the final energy spectra of the 20-fs bunches when laser powers of $P=1$ TW and 2 TW are used. Compared to the results shown in Fig.~\ref{Fig_7_DLA_EbmLenEmit}(d) with $P=0.5$ TW, the maximum energy gain is doubled ($\Delta T_{max}=45$ MeV) by setting $P=2$ TW, as predicted. Since the decelerated electrons cannot continue to meet the QPM condition, their energy loss end near 10 MeV and results in an asymmetric energy distribution with respect to the injection energy $T_0=40$ MeV. However, the strong radial field at $P=2$ TW produces a greater radial force $F_{r}$, which significantly defocuses a fraction of the bunch electrons, as illustrated in the final trace space and electron distributions shown Fig.~\ref{Fig_11_DLA_EbmPwr}(e). This leads to a relatively large final emittance $\epsilon_{N,y}\simeq 18.3 \pi$-mm-mrad. The results indicate that, although the maximum DLA energy gain can be increased by using a laser pulse with higher peak power, the transverse properties of the bunch can be degraded due to the inhibited ion-focusing effect and the increased radial force that causes a greater bunch divergence.

\begin{figure}[tbp]
\centerline{\scalebox{1}{\includegraphics{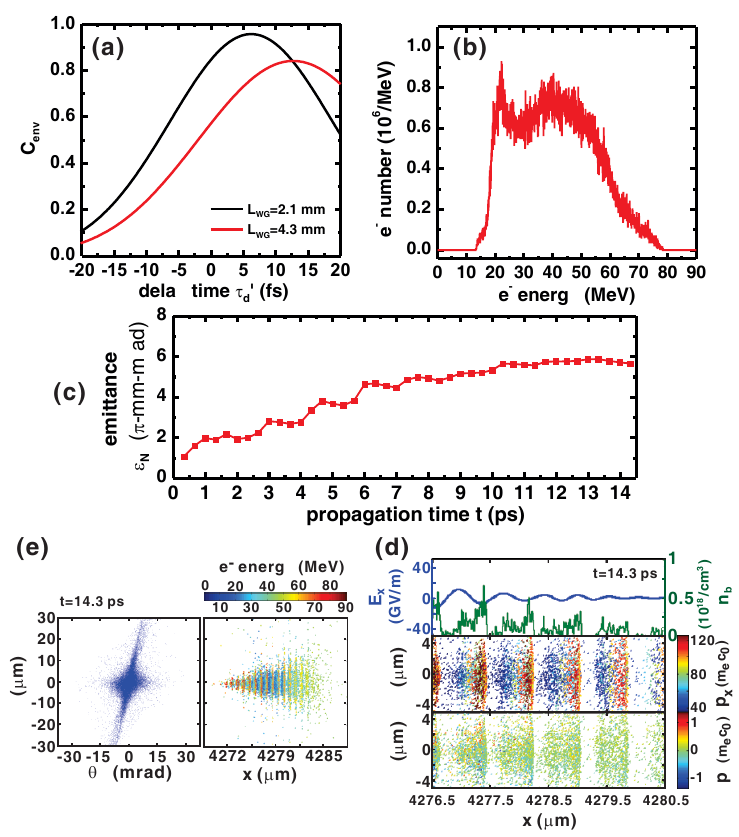}}}
\caption{(a) Dependence of $C_{\mathrm{env}}$ on the delay time $\tau_{d}'$ for waveguide length $L_{wg}=2.1$ mm and 4.3 mm. (b) Final energy spectrum for the bunch with $\tau_{b}=20$ fs with waveguide of $L_{wg}=4.3$ mm and laser power of $P=0.5$ TW. (c) The corresponding emittance $\epsilon_{N,y}$ change with time $t$. Final (d) trace space and electron distributions and (e) on-axis bunch density $n_{b}$ and electron momenta distributions.} 
\label{Fig_12_DLA_EbmLWG}
\end{figure}

A higher maximum energy gain $\Delta T_{max}$ can also be obtained by extending the waveguide length $L_{wg}$. This is studied by adding more density modulation periods into the structure, as illustrated in Fig.~\ref{Fig_1_DLA_PIC_model}(c). A plasma waveguide with $L_{wg}=4.3$ mm, approximately twice the length of the waveguide studied in the previous sections, is studied. The bunch is injected with a delay $\tau_d=0$ with the selected duration $\tau_b=20$ fs, since the ion-focusing effect becomes important in confining the bunch with an extended propagation distance. The factor $C_{\mathrm{env}}$ associated with the averaged field envelope, however, predicts a reduced DLA efficiency with a longer waveguide length $L_{wg}$. The dependence of $C_{\mathrm{env}}$ on the delay $\tau_d'$ when $L_{wg}=2.1$ mm and 4.3 mm is plotted in Fig.~\ref{Fig_12_DLA_EbmLWG}(a). If the bunches are injected with $\tau_b=20$ fs and $\tau_d=0$, the majority of electrons range from $\tau_d'=-10$ fs to 10 fs and have a lowered $C_{\mathrm{env}}$ when $L_{wg}=4.3$ mm. 
In addition, only the trailing electrons with $\tau_d'\sim 8 - 17$ fs can have a relatively high $C_{\mathrm{env}}> 0.8$. As a result, a rapidly decreasing number of electrons up to the maximum gain $\Delta T_{max}\simeq 40$ MeV is present in the final energy spectrum in Fig.~\ref{Fig_12_DLA_EbmLWG}(b). The reduction of $\Delta T_{max}$, compared to the corresponding $\Delta T_{max}=45$ MeV by increasing the laser power to $P=2$ TW is due to the reduced $C_{\mathrm{env}}$ with a longer $L_{wg}$. On the other hand, a small final bunch emittance $\epsilon_{N,y}\simeq 5.7 \pi$-mm-mrad can be obtained when the waveguide length increases, as shown in Fig.~\ref{Fig_12_DLA_EbmLWG}(c), when compared to $\epsilon_{N,y}\simeq 18.3 \pi$-mm-mrad in Fig.~\ref{Fig_10_DLA_EbmSizeEmit}(c) when $P=2$ TW and $L_{wg}=2.1$ mm.  The final trace space and electron distributions in Fig.~\ref{Fig_12_DLA_EbmLWG}(d) show that the ion channel can still confine most of the bunch electrons as they approach the waveguide output, and a relatively small emittance $\epsilon_{N,y}$ can be maintained. However, the axial electron bunching effect cannot not be sustained with an extended $L_{wg}$. As shown in Fig.~\ref{Fig_12_DLA_EbmLWG}(e) by the final on-axis bunch density $n_b$ and the particle axial momentum $P_x$, the decelerated electrons fall behind the energetic electrons in the same $P_x$ modulation period, so that a smoothing in the $n_{b}$ distribution is observed. The results indicate that a higher maximum gain $\Delta T_{max}$ can be realized by use of a longer waveguide. The ion-focusing effect helps maintain the favorable transverse properties of the electron bunch. However, the lower $C_{\mathrm{env}}$ factor, arising from the  walk-off between the laser pulse and the electron bunch, can limit the DLA gain efficiency with an extended waveguide length. This problem can be mitigated by using a longer laser pulse, such that $C_{\mathrm{env}}$ can be increased and its variation with respect to the delay $\tau_d'$ is moderated. In this fashion, a higher energy is needed for the laser pulse in order to retain a high acceleration gradient $E_{x,max}$, which is usually limited by the laser specifications. 
   
\section{\label{sec_Conclu}Conclusion}
  A 3-D PIC model has been developed to simulate QPM of DLA in density-modulated plasma waveguides. Self-consistent solutions for the interactions among the laser pulse, injected electrons, and the backgroun plasma have been obtained, significantly improving the fidelity of the DLA simulations. The model has been applied to simulate the DLA of injected electron bunches in a QPM structure that is designed according to the analytically calculated dephasing lengths. The axial grid size should be chosen as small as computationally reasonable with respect to the laser wavelength ($D_{x}=\lambda/64$ was chosen in this simulation), so that the accuracy of the phase velocity of the laser pulse and the DLA dephasing length, can be maintained in the simulations. Electron bunches, with fundamental properties chosen to match those obtained in typical LWFA experiments, are injected into the structure and accelerated by DLA. A series of studies has been performed by varying the injected bunch length, laser power, and waveguide length to develop an understanding of the DLA performance and sensitivity to those parameters. The effect on the bunch density, trace space, energy spectrum, and emittance has been obtained from the simulations.
  
  When the bunch length is short compared to $\lambda_p/4$ in the low-density region, the choice of the injection delay $\tau_d$ is important for control of the final transverse properties of the bunch. When the injection delay $\tau_d$ is large, a significant divergence of the electron bunch results from the electrostatic force provided by the concentrated  electrons of the background plasma and the defocusing force exerted by the radial field. The collimation of the bunch can be improved by using a smaller injection delay $\tau_d$, in which case a stronger ponderomotive force provided by the laser pulse helps to confine the electrons. However, the maximum energy gain is reduced when a smaller $\tau_d$ is used. When the bunch length becomes closer to $\lambda_p/4$, the ion-focusing effect is enhanced and the final collimation of the bunch can be considerably improved. In this situation, a density modulation of the bunch driven by the radial Lorentz force and the axial momentum modulation can be observed. In the case when the bunch is injected with a large bunch transverse size, comparable to the laser beam diameter, the reduced axial field experienced by the off-axis electrons lowers the acceleration efficiency. On the other hand, the focusing and defocusing of the bunch is enhanced by the stronger radial forces, which contribute to micro-bunch formation. The peak density of the micro-bunches can be approximately 10-fold higher than the peak density of the bunch injected into the waveguide. From those combined results, it can be concluded that the injection of an electron bunch with a long bunch length (close to $\lambda_p/4$, referring to the low-density plasma region) and a small transverse size with respect to the laser pulse diameter is preferred for maintaining the favorable bunch transverse properties in DLA in a plasma waveguide. Under those conditions, the ion-focusing force can effectively collimate the bunch, so that a small emittance can be obtained following the DLA process.     
    
 The maximum energy gain can be increased by increasing the laser power or extending the waveguide length. In the case when a higher power laser pulse is used, the inhibited ion-channel formation and the stronger radial Lorentz force degrade the bunch collimation. The radial force can defocus a large fraction of the electrons in the bunch, even when a relatively long bunch is injected. If the waveguide length is extended to increase the maximum energy gain, the temporal walk-off between the laser pulse and the electron bunch limits the efficiency, such that only the tailing electrons in the bunch can be effectively accelerated to higher energies. From those results we conclude that the optimal DLA requires the use of a moderate laser power to help maintain good transverse properties of the bunch. When the waveguide length is increased, the laser pulse duration must also be increased to mitigate the walk-off effect. However, such longer pulses also requires a greater pulse energy to maintain the the high acceleration gradient.

\section*{Acknowledgements}
This work has been supported by the United States Defense Threat Reduction Agency through contract HDTRA1-11-1-0009 and the National Science Council in Taiwan by grant NSC102-2112-M-008-013. The authors would like to acknowledge the National Center for High-Performance Computing in Taiwan for providing resources under the national project, ``Knowledge Innovation National Grid''.

\newpage

\newpage

\end{document}